\documentclass[12pt,titlepage]{article}

\font\grande=cmr9.5 scaled \magstep4
\font\medio=cmr9.5 scaled \magstep2
\outer\def\beginsection#1\par{\medbreak\bigskip
      \message{#1}\leftline{\bf#1}\nobreak\medskip
\vskip-\parskip
      \noindent}
\usepackage{graphicx} 
\begin{document}
\bibliographystyle {unsrt}

\titlepage

\begin{flushright}
CERN-PH-TH/2010-243
\end{flushright}

\vspace{15mm}
\begin{center}
{\grande Count response model for the CMB spots}\\
\vspace{1.5cm}
 Massimo Giovannini 
 \footnote{Electronic address: massimo.giovannini@cern.ch} \\
\vspace{1cm}
{{\sl Department of Physics, 
Theory Division, CERN, 1211 Geneva 23, Switzerland }}\\
\vspace{0.5cm}
{{\sl INFN, Section of Milan-Bicocca, 20126 Milan, Italy}}
\vspace*{2cm}
\end{center}

\vskip 1cm
\centerline{\medio  Abstract}
The statistics of the curvature quanta generated during a stage of inflationary expansion is used 
to derive a count response model for the large-scale phonons determining, 
in the concordance lore, the warmer and the cooler spots of the large-scale temperature inhomogeneities. The multiplicity distributions for the counting statistics are shown to be generically overdispersed in comparison with conventional Poissonian regressions. The generalized count response model deduced hereunder  
accommodates an excess of correlations in the regime of high multiplicities and prompts 
dedicated analyses with forthcoming data collected by instruments 
of high angular resolution and high sensitivity to temperature variations per pixel. 
\noindent

\vspace{5mm}

\vfill
\newpage
According to the conventional lore of structure formation, 
Cosmic Microwave Background (CMB) observations provide an image of quantum fluctuations 
blown up to the size of the Universe. Curvature perturbations are believed to
originate as quantum fluctuations at least in the framework of the 
standard $\Lambda$CDM paradigm\footnote{In the concordance model (often dubbed $\Lambda$CDM where $\Lambda$ 
stands for the dark energy component and CDM for the dark matter component) single field inflationary models 
and standard thermal history are always assumed implicitly.} where scalar modes of the geometry are the 
sole source of temperature inhomogeneities because of the strict absence of tensor modes.
In this simplified scenario (yet consistent with the three independent cosmological data sets \cite{cmb,bao,sn})
the curvature perturbations are quantized in terms of a collection of scalar phonons and are themselves 
proportional, via the Sachs-Wolfe effect, to the temperature inhomogeneities. 
The latter statement holds for large angular scales (i.e. $\vartheta > 6$ deg) corresponding, in practice, to the region of the 
so-called Sachs-Wolfe plateau, i.e. multipoles $\ell < 30$.
This occurrence is often dubbed by saying that CMB maps at the largest 
scales are a faithful impression of quantum fluctuations as mentioned in the first sentence of this paragraph. For smaller angular scales (i.e., approximately, 
$30\,\mathrm{arcmin} < \vartheta < 6\,\mathrm{deg}$) the curvature perturbations are still the source 
of temperature anisotropies which enter the regime of acoustic oscillations. 
For even larger multipoles (i.e. $\ell > \ell_{\mathrm{S}}$ with $\ell_{\mathrm{S}} \simeq 920$) diffusive (Silk) damping dominates. 

The effective action obeyed by the primordial phonons during a conventional stage of inflationary expansion 
can be written, in terms of ${\mathcal R}$, i.e. the curvature perturbations on comoving orthogonal hypersurfaces which have a gauge-invariant meaning without being necessarily 
connected to curvature perturbations in all the coordinate systems different from the comoving orthogonal one (see, for instance, 
\cite{mg1,pert} and references therein):
\begin{equation}
S= \frac{1}{2} \int d^{3} x \, d\tau z^2 \, \eta^{\alpha\beta} \partial_{\alpha} {\mathcal R} \partial_{\beta} {\mathcal R}, \qquad 
z(\tau) = \frac{a\, \phi'}{{\mathcal H}},
\label{EQ1}
\end{equation}
where $\phi$ denotes the (single) inflaton, $a$ is the scale factor and ${\mathcal H} = (\ln{a})^{\prime}$; the prime denotes
 a derivation with respect to the conformal time coordinate $\tau$. In Eq. (\ref{EQ1}) the geometry is 
assumed to be spatially flat, as suggested by the position of the first acoustic oscillation and by CMB data as a whole (see, e. g.
\cite{cmb}).  In the minimal $\Lambda$CDM scenario (when the non-adiabatic pressure 
fluctuations are strictly vanishing)  the curvature perturbations are approximately constant, i.e. ${\mathcal R}^{\prime} \simeq 0$ for wavelengths shorter than Hubble radius  at each corresponding time (see, for instance, \cite{mg2}).  In particular, at photon decoupling, the intensity fluctuations of the radiation field (i.e. the warmer and the cooler regions in the CMB sky) 
are given, in real space, as 
\begin{equation}
\Delta_{\mathrm{I}}(\vec{x},\tau_{\mathrm{dec}}) \simeq - \frac{{\mathcal R}(\vec{x},\tau_{\mathrm{dec}})}{5},
\label{EQ1A}
\end{equation}
under the approximation of sudden decoupling  (i.e. assuming that the visibility function is a narrow 
Gaussian centered at $\tau_{\mathrm{dec}})$.  Following the tenets of canonical quantization, the field variables can be promoted to quantum mechanical operators obeying equal-time commutation relations, i.e. 
\begin{equation}
\hat{{\mathcal R}}(\vec{x},\tau) = \frac{1}{(2\pi)^{3/2}\,\sqrt{V}} \sum_{\vec{k}} \hat{{\mathcal R}}_{\vec{k}}\, e^{- i \vec{k}\cdot\vec{x}}, \qquad 
\hat{{\mathcal R}}_{\vec{k}} = \frac{\hat{a}_{\vec{k}} + \hat{a}^{\dagger}_{-\vec{k}}}{z(\tau) \sqrt{2 k}},
\label{EQ2}
\end{equation}
where $\hat{a}_{\vec{k}}$ and $\hat{a}^{\dagger}_{p}$ are the annihilation and creation operators of curvature inhomogeneities 
and $[\hat{a}_{\vec{k}}, \hat{a}^{\dagger}_{\vec{p}}] = \delta_{\vec{k},\, \vec{p}}$; per each Fourier mode 
the Hamiltonian operator describing the quantized curvature perturbations can be written as:
\begin{equation}
\hat{H}_{\vec{k}} = 2 k \,\, {\mathcal K}_{0}(\vec{k})  +2 \biggl[ \lambda^{*}(\tau) {\mathcal K}_{-}(\vec{k}) + \lambda(\tau) {\mathcal K}_{+}(\vec{k})\biggr],
\label{EQ3}
\end{equation}
where $2 \lambda = i z^{\prime}/z$; the operators ${\mathcal K}_{\pm}(\vec{k})$ and ${\mathcal K}_{0}(\vec{k})$ obey the commutation relations of the $SU(1,1)$ Lie algebra \cite{pere,schum}:
\begin{equation}
{\mathcal K}_{+}(\vec{k}) = \hat{a}_{\vec{k}}^{\dagger} \,\hat{a}_{-\vec{k}}^{\dagger},\qquad {\mathcal K}_{-}(\vec{k}) = \hat{a}_{\vec{k}}\, \hat{a}_{-\vec{k}},
\qquad {\mathcal K}_{0}(\vec{k}) = \frac{1}{2}\biggl[ \hat{a}_{\vec{k}}^{\dagger}\, \hat{a}_{\vec{k}} + \hat{a}_{-\vec{k}} \,\hat{a}_{-\vec{k}}^{\dagger}\biggr].
\label{EQ4}
\end{equation}
From Eq. (\ref{EQ3}) the multiparticle final state is given by 
\begin{equation}
|\Psi_{\vec{k}} \rangle = \Xi(\varphi_{k}) \, \Sigma(\zeta_{k}) |0_{\vec{k}} \, 0_{-\vec{k}}\rangle, \qquad |\Psi \rangle = \prod_{\vec{k}} |\Psi_{\vec{k}}\rangle,
\label{EQ5}
\end{equation}
where the two unitary operators $ \Xi(\varphi_{k})$ and $\Sigma(\zeta_{k})$ are  
\begin{equation}
 \Xi(\varphi_{k}) = \exp{[ - 2 i \varphi_{k}\, {\mathcal K}_{0}(\vec{k})]}, \qquad \Sigma(\zeta_{k}) = \exp{[\zeta_{k}^{*}\, {\mathcal K}_{-}(\vec{k}) - \zeta_{k}\, {\mathcal K}_{+}(\vec{k})]},
\label{EQ6}
\end{equation}
with $\zeta_{k} = r_{k} e^{i \gamma_{k}}$ and $\alpha_{k} = (2\varphi_{k} -\gamma_{k})$; the time evolution of the 
variables $r_{k}(\tau)$, $\varphi_{k}(\tau)$  and $\alpha_{k}(\tau)$ is given by 
\begin{equation}
 r_{k}^{\prime} = 2 i \lambda \cos{\alpha_{k}},\qquad \varphi_{k}^{\prime} = k  -  2 i \lambda \tanh{r_{k}} \sin{\alpha_{k}},\qquad \alpha_{k}^{\prime} = 2 k - 4 i \lambda \frac{\sin{\alpha_{k}}}{\tanh{2 r_{k}}}.
\label{EQ7}
\end{equation}
In Eq. (\ref{EQ5}) the initial state is the vacuum. While the latter assumption can be relaxed, it is usually 
invoked by tacitly assuming that the total number of inflationary efolds exceeds the minimal amount required 
to solve the standard problems of big-bang cosmology (see, e.g. \cite{mg1}, first reference). 

The customary exercise would now be to compute 
the power spectrum in terms of the parameters of the underlying inflationary model encoded in the function $z(\tau)$ 
defined in Eq. (\ref{EQ1}). A qualitatively different class of questions concerns instead the determination of the 
multiplicity distribution of the curvature quanta.  Owing to the $SU(1,1)$ group structure of Eq. (\ref{EQ4}), the squeezing 
operator of $\Sigma(\zeta_{k})$ defined in Eq. (\ref{EQ5}) can be factorized as $\Sigma(\zeta_{k}) = {\mathcal A}_{+}(\zeta_{k}) \, {\mathcal A}_{0}(\zeta_{k}) {\mathcal A}_{-}(\zeta_{k})$ 
where ${\mathcal A}_{0}(\zeta_{k}) = \exp{[ - 2 \ln{(\cosh{r_{k}})} {\mathcal K}_{0}(\vec{k})]}$ and ${\mathcal A}_{\pm}(\zeta_{k})  = \exp{[\mp e^{\pm i \gamma_{k}} \tanh{r_{k}} \, {\mathcal K}_{\pm}(\vec{k})]}$. From Eq. (\ref{EQ5}) the form of the 
density matrix relevant for the forthcoming discussions is 
\begin{equation}
\hat{\rho}_{\vec{k}} = \frac{1}{\cosh^2{r_{k}}} \sum_{n_{\vec{k}}=0}^{\infty} \sum_{m_{\vec{k}}=0}^{\infty} e^{- i \alpha_{k}(n_{\vec{k}} - m_{\vec{k}})} 
(\tanh{r_{k}})^{n_{\vec{k}} + m_{\vec{k}}} |n_{\vec{k}}\,\, n_{- \vec{k}}\rangle \langle m_{-\vec{k}}\,\,m_{\vec{k}}|,
\label{QM2a}
\end{equation}
whose diagonal elements define the multiplicity distribution: 
\begin{equation}
P_{\{n_{\vec{k}}\}} = \prod_{\vec{k}} P_{n_{\vec{k}}} , \qquad P_{n_{\vec{k}}}(\overline{n}_{\vec{k}})= 
\frac{\overline{n}_{\vec{k}}^{n_{\vec{k}}}}{( 1 + \overline{n}_{\vec{k}})^{n_{\vec{k}} + 1 }},
\label{QM2}
\end{equation} 
accounting for the way curvature quanta of each Fourier mode (i.e. $n_{\vec{k}}$) 
 are distributed as a function of their mean value per each Fourier mode (i.e.  $\overline{n}_{\vec{k}} = \sinh^2{r_{k}}$).  
 Equation (\ref{QM2}) is a  Bose-Einstein distribution but the average number 
of curvature quanta $\overline{n}_{\vec{k}}$ has no relation with the standard Bose-Einstein occupation number. 
The same situation occurs usually in quantum optics for chaotic (i.e. white) light 
\cite{mandel,loudon} where photons distributed as in Eq. (\ref{QM2}) for each mode of the radiation field 
can be produced by sources in which atoms are kept at an excitation level higher than that in thermal equilibrium.
The way the off-diagonal elements of Eq. (\ref{QM2a}) behave is dictated by the phases  
$\alpha_{k}$. While it is not strictly essential to get rid of the off-diagonal elements for the forthcoming 
arguments, it is nonetheless plausible that, by averaging over $\alpha_{k}$ Eq. (\ref{QM2a}), 
 the density matrix can be reduced 
 (i.e. $\hat{\rho}^{\mathrm{red}}_{\vec{k}} = \frac{1}{2\pi} \int_{0}^{2\pi} d \alpha_{k} \hat{\rho}_{\vec{k}}$) by only keeping 
 the diagonal terms. 
The density operator can be written, with the conventional
shorthand notation,
\begin{equation}
\hat{\rho} = \sum_{\{n_{\vec{k}}\}} P_{\{n_{\vec{k}}\}}  \, | \{n_{\vec{k}}\} \rangle \langle  \{n_{\vec{k}}\}|,\qquad 
| \{n_{\vec{k}}\} \rangle  =|n_{\vec{k_{1}}} \rangle |n_{\vec{k_{2}}} \rangle|n_{\vec{k_{3}}} \rangle...
\label{QM1}
\end{equation}
where, $P_{\{n_{\vec{k}}\}}$ is given by Eq. (\ref{QM2}) and the ellipses stand for the direct product of 
all the modes of the field.  According to Eq. (\ref{EQ1A}) (and in the hypothesis that curvature quanta 
are the sole source of temperature inhomogeneities), the distribution $P(n)$ of the {\em total} number of phonons 
$n = \sum_{\vec{k}} n_{\vec{k}}$ must reflect the distribution of the warmer and cooler regions. If there are supplementary 
sources of temperature inhomogeneities (e. g. tensor modes of the geometry) 
the present discussion can be appropriately modified. The multiplicity distribution $P(n)$ accounts for the way the total number 
of phonons $n$ is distributed as a function of its mean value; $P(n)$ 
can be very different from $P_{\{n_{\vec{k}}\}}$.  Denoting with $p(\{n\})$ the joint probability distribution of the set of phonon occupation numbers $\{n\}$ of the field, we shall have that 
\begin{equation}
p(\{n\}) = \prod_{\vec{k}}\frac{1}{(1 + \overline{n}_{\vec{k}}) ( 1 + 1/\overline{n}_{\vec{k}})^{n_{\vec{k}}}}.
\label{QM4}
\end{equation}
For any mode for which 
$\overline{n}_{\vec{k}} =0$, the corresponding factor must be interpreted  as $\delta_{n_{\vec{k}} \, 0}$. In the following 
we shall suppose, quite generally, that only a subset consisting of 
$\epsilon$ modes of the field is actually occupied and we shall restrict the attention to this subset of modes. 
If $n$ is the total number of phonons and $P(n)$ is the multiplicity distribution of $n$, then 
$P(m) = \sum_{\{n\}} p(\{n\}) \delta_{m \, n}$. In quantum optics an analog of the multiplicity distribution $P(m)$ 
describes the statistical properties of (unpolarized) chaotic light beams \cite{mandel}.  
The evaluation of $P(m)$ can be in general difficult but 
it becomes easy in the physical case when the average occupation number $\overline{n}_{\vec{k}}$ of all the $\epsilon$ occupied modes become equal\footnote{In the case of a thermal light beam  which is either fully polarized or fully unpolarized the use 
of a rectangular spectral density is an excellent approximation in the derivation of the 
photocounting statistics which is also experimentally accessible \cite{mandel}.}; in this case  from Eq. (\ref{QM1}) and (\ref{QM4}) the joint probability distribution of the occupied modes becomes:
\begin{equation}
p(\{n\}) = \frac{1}{(1 +  \overline{n}/\epsilon)^{\epsilon} ( 1 + \epsilon/\overline{n})^{n}}, \qquad \overline{n} = \sum_{\vec{k}} \overline{n}_{\vec{k}} = \epsilon \overline{n}_{\vec{k}}.
\label{QM7}
\end{equation}
Every non-vanishing term in the summation $P(m) = \sum_{\{n\}} p(\{n\}) \delta_{m \, n}$ has  the same 
value and the required probability 
$P(m)$ is simply $p(\{n\})$  given by Eq. (\ref{QM7}) multiplied by a well known combinatorial factor accounting 
for the way the $n$ phonons are distributed among the $\epsilon$ modes:
\begin{equation}
P_{n}(\overline{n},\epsilon)=  \frac{\Gamma( n + \epsilon)}{\Gamma(\epsilon) \Gamma(n + 1)} \biggl(\frac{\overline{n}}{\overline{n} + \epsilon}\biggr)^{n} \biggl(\frac{\epsilon}{\overline{n} + \epsilon}\biggr)^{\epsilon}.
\label{QM8}
\end{equation}
The cumulant generating function \cite{barucha} associated with Eq. (\ref{QM8}) is given by 
 \begin{equation}
{\mathcal C}(s, \overline{n}, \epsilon) = \ln{\biggl[ \sum_{n} s^{n} P_{n}(\overline{n},\epsilon)\biggr]}
 = - \epsilon \ln{\biggl[ 1 + ( 1 - s) \frac{\overline{n}}{\epsilon}\biggr]}.
 \label{QM9}
 \end{equation}
Equations (\ref{QM8}) and (\ref{QM9}) define the count response model implied by the physical 
nature of the source which are the (quantized) curvature perturbations. From the cumulant generating function 
all the cumulant moments can be obtained to an arbitrary order but they are all function of $\overline{n}$ 
and of $\epsilon$. The variance $D^2 = \langle n^2\rangle - \langle n \rangle^2$ is then given 
by $D^2 =  \overline{n} + \overline{n}^2/\epsilon$. In the limit $\epsilon \to 1$ the Bose-Einstein
distribution is recovered; for generic $\epsilon$ the count response model defined by Eqs. (\ref{QM8}) and (\ref{QM9}) 
falls into the class of negative binomial regressions which arise, in rather general terms, in all 
those discrete counts where correlations lead to asymmetric distributions with a degree of correlation larger than 
in conventional Poissonian counting (see \cite{regr} for an introduction 
to statistical models of count response data). 

The count response model of Eqs. (\ref{QM8})  and (\ref{QM9}) has been deduced in a top-down approach 
by looking at the statistical properties of the counting distribution of primordial phonons 
as they arise in the minimal $\Lambda$CDM model. A complementary avenue will now be taken
with the purpose of deriving the concept of multiplicity distribution in a bottom-up perspective. 
To study the multiplicity distribution of the spots we can, for instance, fix a threshold in the brightness perturbations such as 
\begin{equation}
|\Delta_{\mathrm{I}}^{(\mathrm{min})}| = 26.865 \biggl(\frac{{\mathcal A}_{{\mathcal R}}}{2.43\times 10^{-9}}\biggr)^{1/2} \biggl(\frac{T_{\gamma 0}}{2.725\, \mathrm{K}}\biggr)\,\, \mu\mathrm{K},
\label{CONC2}
\end{equation}
where ${\mathcal A}_{{\mathcal R}}$ is the amplitude of the power spectrum of curvature perturbations 
following from the WMAP 7 data at the conventional pivot scale $k_{\mathrm{p}} =0.002\, \mathrm{Mpc}^{-1}$. The value of Eq. (\ref{CONC2}) 
comes from the large-scale ($\vartheta> \,6$deg) plateau; different and more refined ways of fixing the threshold 
can be suggested but this aspect is immaterial for the forthcoming considerations
\footnote{We are here focussing on hot spots but the same 
discussion can be conducted for cooler regions by considering the distribution of spots below a given (progressively decreasing) threshold. This the meaning of the absolute value in Eq. (\ref{CONC2}).}.  Given the threshold (\ref{CONC2}) 
we can therefore ask (or predict)  how many hot (or cold) spots 
are present in different angular intervals starting from a $\vartheta_{\mathrm{min}}$ 
(connected, for instance, with the resolution of the instrument) up to a $\vartheta_{\mathrm{max}}$.
The resolution\footnote{The WMAP experiment in his five channels has a resolution which varies between about $1$ deg (for the 
low frequency channels) to less than $0.2$ deg for the high-frequency channels. 
Th Planck explorer experiment, in his nine frequency channels has a resolution 
which varies between few arcminutes (in the high frequency instrument) and about 
half a degree of the low frequency channels.}  will also determine ultimately the nature of the partition 
and the number of classes of the histogram. The same procedure will then to be 
used for different temperature thresholds. 
It is plausible to assume, in a bottom-up approach, that the distribution of spots 
in excess with respect to a given (progressively increasing) threshold
is given by a real Gaussian variable $f(\vartheta)$. In a Poissonian count response model the probability that  one spot is found between $\vartheta$ and $\vartheta + d \vartheta$ 
will be $P(1, \vartheta, d\vartheta) = \mu {\mathcal Q}(\vartheta) \,d\vartheta$ where $\mu$ is a constant and \cite{barucha}
\begin{equation}
{\mathcal Q}(\vartheta) = \frac{1}{\delta} \int_{\vartheta - \delta/2}^{\vartheta + \delta/2} f^2(\vartheta) \, d\vartheta.
\label{BU1}
\end{equation}
The probability of finding $n$ spots in the interval $[\vartheta,\, \vartheta + \Delta\vartheta]$ is a Poisson distribution in $n$ 
\begin{equation}
P(n, \theta,\vartheta) = \frac{1}{n!} \biggl[\mu \int_{\vartheta}^{\vartheta + \Delta\vartheta} {\mathcal Q}(\vartheta) d\vartheta\biggr]^{n} \exp{\biggl[ - \mu 
\int_{\vartheta}^{\vartheta + \Delta\vartheta} {\mathcal Q}(\vartheta^{\prime}) \,\,d\vartheta^{\prime}\,\,\biggr]}.
\label{BU2}
\end{equation}
It would be debatable to conclude that the multiplicity distribution is Poissonian  
since what is potentially measurable is not $P(n,\vartheta,\Delta\vartheta)$ but rather $p_{n}(\Delta\vartheta)= \langle P(n,\vartheta,\Delta\vartheta)\rangle$, 
i.e. the distribution $P(n,\vartheta,\Delta\vartheta)$ expressed as a function of $x= \int_{\vartheta}^{\vartheta + \Delta\vartheta} {\mathcal Q}(\vartheta') \, d\vartheta'$
and averaged over the ensemble of $x$.  The mean and the variance of $x$ can be physically estimated as $\langle x\rangle = \overline{{\mathcal Q}}\, \Delta\vartheta$
and as $\langle x^2 \rangle - \langle x\rangle^2 =  \overline{{\mathcal Q}}^2\, (\Delta\vartheta)\, \vartheta_{\mathrm{c}}$ where $\vartheta_{\mathrm{c}}$ is a 
typical scale possibly related to the angular resolution to some other coarsening angle. The question is therefore if it exists a 
probability distribution of $x$ (containing at least two parameters) and reproducing the results already obtained within our top-down approach.  The answer to the latter question is provided by the Gamma distribution \cite{barucha}
\begin{equation}
p_{\Gamma}(\lambda,\,\nu;\,x) \, d x = (\lambda x)^{\nu -1} \frac{e^{- \lambda x}}{\Gamma(\nu)} d (\lambda x), \qquad \lambda = \frac{1}{\overline{Q}\,\vartheta_{c}},\qquad 
\nu = \frac{\Delta\vartheta}{\vartheta_{\mathrm{c}}}.
\label{BU3}
\end{equation}
The distribution of Eq. (\ref{BU2}) must then be averaged over the ensemble of $x$ 
\begin{equation}
p_{n}(\Delta\vartheta) = \frac{1}{n!}  \int_{0}^{\infty}p_{\Gamma}\biggl(\frac{1}{\overline{Q}\vartheta_{\mathrm{c}}}, \frac{\Delta\vartheta}{\vartheta_{\mathrm{c}}}; x\biggr)
e^{- \mu x} \, (\mu x)^{n} \, d x.
\label{BU4}
\end{equation}
The explicit result of the integral indicated in Eq. (\ref{BU4}) is exactly given by Eq. (\ref{QM8}) with $\overline{n} = \mu \overline{{\mathcal Q}} \Delta\vartheta$ and $\epsilon =( \Delta\vartheta)/\vartheta_{\mathrm{c}}$. 
The  Poissonian counting is recovered from Eq. (\ref{QM9}): in the limit $\epsilon \to \infty$ the variance tends 
to the mean value (i.e. $D^2 \to \overline{n}$)  and the cumulant generating function of Eq. (\ref{QM9}) 
tends to the Poissonian limit, i.e. ${\mathcal C}(s, \overline{n}, \epsilon)\to \overline{n}(s-1)$. These two occurrences 
are sufficient to infer that, for $\epsilon \to \infty$, $P_{n}(\overline{n},\epsilon) \to \overline{n}^{n} \,\exp{[- \overline{n}]}/n!$.
A posteriori we can say that the negative binomial counting deduced in Eq. (\ref{QM8}) (and, indirectly, in Eq. (\ref{BU4})) is, at once, 
more general and more physical than the Poissonian counting. 

The probability of finding $n$ spots in a given angular interval must depend on the 
threshold. By varying the threshold of Eq. (\ref{CONC2}) as 
$|\Delta_{\mathrm{I}}^{(\mathrm{min})}|\to |\Delta_{\mathrm{I}}^{(\mathrm{min})}| \eta$ (where $\eta \geq 1$)
the values of $\overline{n}$ and $\epsilon$ will depend upon $\eta$ (in particular we can intuitively 
expect that $\overline{n}$ will decrease with $\eta$).  The probability generating 
function ${\mathcal P}$ of the  negative binomial distribution satisfies 
\begin{eqnarray}
&&\frac{d {\mathcal P}}{d \eta} =  - {\mathcal G}({\mathcal P},\eta),
\label{CONC6}\\
&& {\mathcal G}({\mathcal P},\eta)= -\frac{1}{\epsilon} \frac{d\epsilon}{d\eta} {\mathcal P} \ln{{\mathcal P}} +
\frac{\epsilon^2}{\overline{n}} \frac{d}{d\eta} \biggl(\frac{\overline{n}}{\epsilon}\biggr)
{\mathcal P}[ 1 - {\mathcal P}^{1/\epsilon}],
\label{CONC7}
\end{eqnarray}
which has the form of a reverse Kolmogorov equation and where ${\mathcal P}(\overline{n},\epsilon) = \sum_{n=0}^{\infty} P_{n}(\overline{n},\epsilon)$.
It is tempting to interpret $\eta$ as a continuous evolution parameter of an appropriate 
branching process \cite{barucha}. In this case we should probably impose the boundary 
conditions ${\mathcal G}(1, \eta) =0$ and ${\mathcal G}({\mathcal P}, \eta) =0$ (for ${\mathcal P} \to 0$) and ask that 
${\mathcal G}(P,\eta)$ factorizes as the product of a function of ${\mathcal P}$ and of a function 
of $\eta$: in this way the branching process will be stationary in $\eta$ and Eqs. (\ref{CONC6})--(\ref{CONC7}) imply that one (or both) the conditions are satisfied 
\begin{equation}
\frac{1}{\epsilon} \frac{d \epsilon}{d\eta} = c_{1} \frac{\epsilon^2}{\overline{n}} \frac{d}{d\eta} 
\biggl(\frac{\overline{n}}{\epsilon}\biggr),\qquad {\mathcal P} \ln{{\mathcal P}} = c_{2} {\mathcal P} ( 1 - {\mathcal P}^{1/\epsilon})
\label{CONC8}
\end{equation}
where $c_{1}$ and $c_{2}$ are two numerical constants. This means that, assuming 
that the stochastic processes is stationary the $\eta$-dependence of $\overline{n}$ 
determines also the $\eta$-dependence of $\epsilon$. For instance, from the first relation 
of Eq. (\ref{CONC8}) we have that $1/\epsilon(\eta) + \ln{\epsilon(\eta)} = a_{1} + b_{1} \ln{\overline{n}(\eta)}$.

The modest but novel purpose of the present paper could be summarized by saying that 
to count spots in a physically meaningful way we have to understand which is the appropriate count response model.  It has then been suggested how the multiplicity distribution of the total number of phonons 
is connected to the multiplicity distribution of the CMB spots in different angular intervals and for different temperature thresholds.
These physical considerations pin down a specific count response model which is overdispersed in comparison with a naive Poissonian counting.  While the present considerations can be generalized to related frameworks it seems interesting to pursue dedicated analyses aimed at measuring the multiplicity distribution of CMB spots. The experimental scrutiny will therefore have to 
infer the likely values of $\epsilon$ and $\overline{n}$ for different thresholds 
$\eta$ and for different angular intervals (and, presumably, also for different frequency channels of the instrument). 
The latter program assumes that the resulting CMB signal used to extract the multiplicity distribution will be free of the contamination of foreground (e.g. point-like) sources, a potentially severe problem whose discussion and is beyond the scopes of the present paper.

\end{document}